\documentclass[fleqn,10pt]{wlscirep}
\usepackage{graphicx}
\usepackage{hyperref}
\title{Resonant X-ray emission with a standing wave excitation}

\author[1,*]{Kari O. Ruotsalainen}
\affil[1]{Department of Physics, P.~O.~Box 64, FI-00014 University of Helsinki, Finland}
\author[1]{Ari-Pekka Honkanen}
\author[2]{Stephen P. Collins}
\affil[2]{Diamond Light Source, Didcot, Oxfordshire, OX11 0DE}
\author[3]{Giulio Monaco}
\affil[3]{Physics Department, University of Trento,Via Sommarive 14, 38123 Povo (TN), Italy}
\author[4]{Marco Moretti Sala}
\affil[4]{European Synchrotron Radiation Facility, 71 avenue des Martyrs, 38000 Grenoble, France}
\author[4]{Michael Krisch}
\author[1]{Keijo H{\"a}m{\"a}l{\"a}inen}
\author[1]{Mikko Hakala}
\author[1]{Simo Huotari}

\affil[*]{corresponding author: kari.ruotsalainen@helsinki.fi}

\begin{abstract}
The Borrmann effect is the anomalous transmission of x-rays in perfect crystals under diffraction
conditions. It arises from the interference of the incident and diffracted waves, which 
creates a standing wave with nodes at strongly absorbing atoms. Dipolar absorption of
x-rays is thus diminished,  which makes the crystal nearly transparent for certain 
x-ray wave vectors. Indeed, a relative enhancement of electric quadrupole absorption via the 
Borrmann effect has been demonstrated recently. Here we show that the Borrmann effect has a significantly larger
impact on resonant x-ray emission than is observable in x-ray absorption.
Emission from a dipole
forbidden intermediate state may even dominate the corresponding x-ray spectra.
Our work extends the domain of x-ray standing wave 
methods to resonant x-ray emission spectroscopy and provides means for novel 
spectroscopic experiments in d- and f-electron systems. 
\end{abstract}
\begin{document}

\flushbottom
\maketitle

%
\thispagestyle{empty}

\section*{Introduction}
The pursuit of understanding the microscopic origins of the properties
of bulk matter often relies on spectroscopy and scattering of x-ray
photons. X-ray absorption and emission spectroscopies probe the
unoccupied and occupied electronic states,
respectively \cite{degroot01,glatzel05}. 
Magnetic circular and linear as well as natural circular dichroism in x-ray 
absorption spectra give access to ground state spin and angular momentum 
expectation values, the spin-orbit coupling constant 
and structural information \cite{carra90,thole92,vanderlaan99,lucilla98}.
Resonant x-ray emission and resonant inelastic x-ray scattering spectroscopies yield
element, spin and orbital selective electronic structure
information and probe valence electron excitations \cite{kotani01}.
A much utilized advantage of resonant x-ray emission spectroscopy
over traditional x-ray absorption spectroscopy is the ability 
to resolve life time broadening limited features in the x-ray absorption 
spectrum \cite{hamalainen91,kotani01}. It can be stated that the 
introduction of resonant x-ray spectroscopies turned over the notion 
that the core hole lifetime is a fundamental limit to the energy resolution 
obtainable in x-ray absorption spectra.

On the other hand, x-ray standing wave techniques provide an attractive
approach to achieve site sensitive spectroscopic
information \cite{vartanyants01}. For bulk solids, applications of
standing wave methods have included e.g. locating impurity sites and
solving the phase problem of x-ray
diffraction \cite{batterman64_2,batterman69,bedzyk85,vartanyants01}.
There also are several varieties of standing wave spectroscopies
applicable to thin films, multilayers and surfaces, most impressive
demonstrations of which include the determination of adsorption sites
on surfaces \cite{cowan80,bedzyk85_2,vartanyants01,woodruff05}.
X-ray spectroscopies and standing wave fields 
can thus be joint in very fruitful combinations that yield novel 
information on materials relevant for physics and chemistry.

Under the diffraction condition, the incident and diffracted waves in
a crystal form a coherent superposition,
with a standing wave along the scattering vector \cite{batterman64}.
The relative amplitudes and phases of the waves depend on the
deviation from the Bragg angle. In the two-beam case with
monochromatic linearly polarised plane waves, the Maxwell equations
have two solutions per polarization state with respect to the
scattering plane (parallel $\pi$, perpendicular $\sigma$).  One of the
solutions (so-called $\alpha$-branch) has nodes on atomic sites, and
the other ($\beta$-branch) corresponds to antinodes at those locations.
The absorption owing to the electric dipole (E1) term depends on 
the field amplitude at the atomic site and thus the attenuation 
of the $\alpha$-branch is diminished while 
the $\beta$-branch, in turn, is absorbed
rapidly \cite{batterman64,pettifer08}. In the Laue (transmission) geometry, exciting the $\alpha$-branch
gives rise to a well known manifestation of dynamical diffraction:
the anomalous transmission of x rays known as the Borrmann effect \cite{batterman64}.

There are fascinating ways to use the Borrmann effect, e.g., 
in x-ray absorption spectroscopy. For instance, Pettifer \textit{et al.} 
demonstrated using the Borrmann effect a very large relative enhancements 
of electric quadrupole resonances at the Gd L edges in Gd$_3$Ga$_5$O$_{12}$, 
where the normally weak quadrupole-allowed pre-edges almost reached the 
intensity of the dipolar main edges at the temperature of 10~K 
\cite{pettifer08}. Identification of the transition multipolarity is of 
utmost importance in the interpretation of magnetic x-ray dichroic 
spectroscopies and their study provided unambiguous proof of the quadrupolar
barely visible pre-edge at the Gd L$_3$ edge. 
This solved a long-standing issue of the nature of the pre-edge resonance,  
 which had been a much discussed topic in the context of x-ray
circular magnetic dichroism of rare earth compounds \cite{carra91,lang92,wang93,lang95,giorgetti95}.

One way to gain insight to their dipolar and quadrupolar 
contributions to x-ray spectra would be to utilise their
different angular dependencies. However, such an analysis may 
require support from parameter-dependent crystal or ligand field 
calculations.  Moreover, utilising the angular
dependence of the multipoles is complicated in certain systems where they
have similar angular behaviour \cite{carra90,krisch95}. The Gd$^{3+}$ ion in GGG is one well
known case of where alternative methods were needed to understand the 
absorption edge structure. To this end, 
Krisch \textit{et al.} applied resonant x-ray
emission spectroscopy to GGG and provided strong evidence for the  
quadrupolar nature on the pre-edge feature at the Gd L$_3$ edge \cite{krisch95}.
They measured the intensity of the Gd L$\alpha_1$ emission line while
tuning the incident x-ray energy across the Gd L$_3$ edge,
revealing the quadrupole-excited state via its fingerprint in the emission spectrum.

In this work, we combine for the first time 
resonant x-ray emission spectroscopy and
the Borrmann effect, using the Gd L$_3$ resonance in GGG.
We exploit the effective transparency of the crystal to 
the anomalously transmitted x rays in a novel experiment 
that allows us to enhance the quadrupole-allowed states in the
resonant x-ray emission spectra. We demonstrate a tenfold increase in
the relative weight of x-ray emission from a dipole-forbidden
intermediate state at the Gd L$_3$ edge 
when excited by anomalously transmitted light. 
Our results thus show that high resolution
x-ray spectroscopy can be successfully augmented with 
standing wave techniques with promising applications.

\begin{figure}
\centering
 \includegraphics[hiresbb]{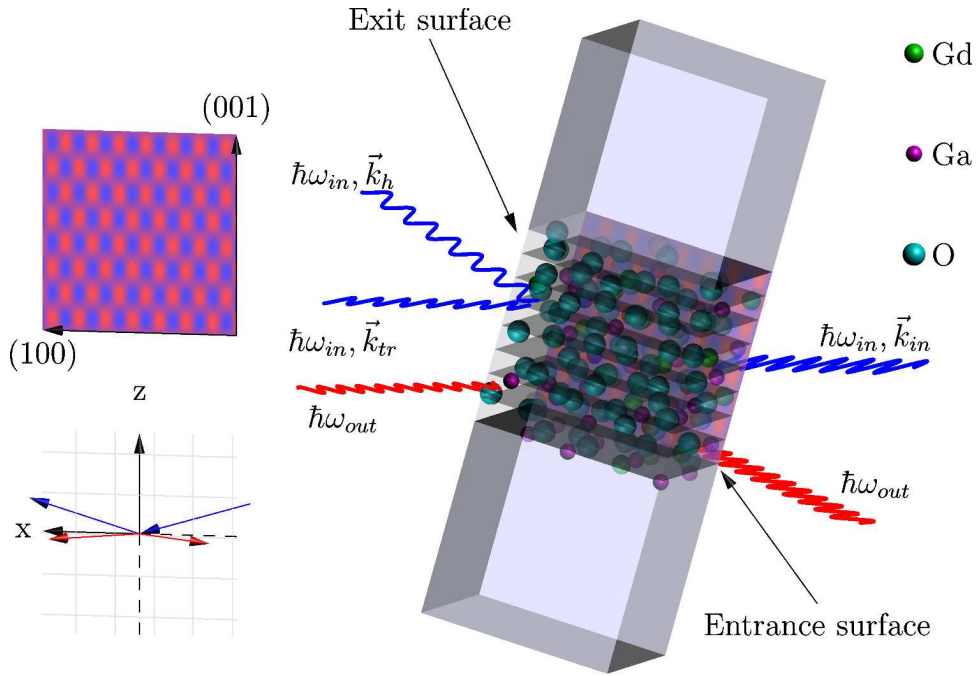}%
 \caption{\label{ref:exp} Illustration of the experiment. Top left : The wavefield pattern 
 with respect to the (100) and (001) lattice directions in the conventional cubic unit cell of GGG.
 The standing wave is formed along the (001) direction.   
 Bottom left: The diffraction and fluorescence observation geometries.
 The xz-plane is the vertical scattering plane, in which the anomalously transmitted beams are observed. 
 The spectrometer observes fluorescence  in the xy plane, and is rotated about the z
axis for observing the emitted radiation from the entrance and exit surfaces of the sample. 
Right: The crystal structure of GGG and the (008) lattice planes are visualized in the center. 
Blue wavy lines represent the incident, transmitted and diffracted photons with E$_\mathrm{in}$=$\hbar\omega_{\mathrm{in}}$ and
wave vectors $\vec{k}_{in}$, $\vec{k}_{tr}$, and $\vec{k}_h$.
The incident beam impinges on the entrance surface (right-hand side of crystal). In anomalous transmission the 
energy carried by the incident beam flows along the diffracting planes and two beams emerge from the exit surface (left-hand side of the crystal).
The redshifted fluorescence photons (E$_\mathrm{out}$=$\hbar\omega_\mathrm{out}$) are represented by the red lines. }
 \end{figure}

\section*{Results}

\begin{figure}[ht]
\centering
 \includegraphics[width=1\textwidth]{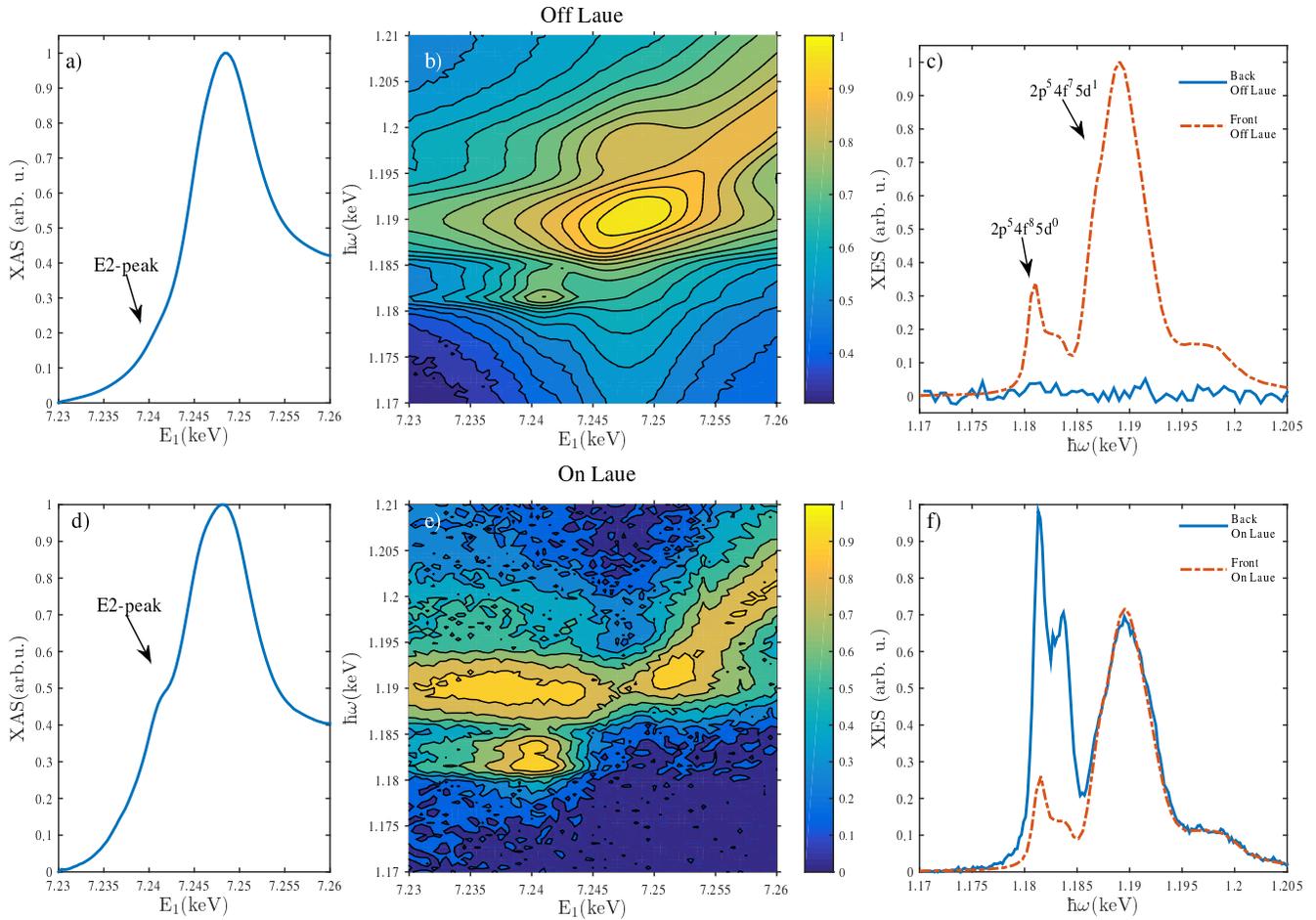}%
 \caption{\label{fig:megaplot}  a) Entrance surface off-Laue total fluorescence yield x-ray absorption spectrum of the Gd L$_3$ 
edge, b) 
the associated entrance surface resonant x-ray emission map on 
a logarithmic scale to highlight the weak emission lines, 
c) the corresponding resonant x-ray emission spectrum excited 
at the pre-edge ($E_\mathrm{in}$=7.2405 keV). The solid and dashed lines denote spectra recorded from 
the entrance and exit surfaces of the crystal, respectively. d) The transmission x-ray absorption spectrum under the diffraction condition. e) 
Exit-surface resonant x-ray emission map on a logarithmic scale. Note the large 
difference in comparison with Fig. 2b). The low intensity regions of the map have been smoothed 
for visual clarity. f) The resonant x-ray emission spectrum under the diffraction condition for $E_\mathrm{in}$=7.2405 keV. X-ray emission from the 
quadrupole-allowed 4f$^8$5d$^0$ intermediate state gains more weight than the dipole-allowed 4f$^7$5d$^1$ one.  }
 \end{figure}

We performed a resonant x-ray emission experiment on a GGG single crystal 
to study the emission spectrum under the Borrmann effect. 
We tuned the incident x-ray energy $E_\mathrm{in}$  
across the Gd L$_3$ edge and  for each energy 
measured the resonant Gd L$\alpha_1$ emission (spectrometer recording intensity at a 
scattered photon energy $E_\mathrm{out}$).
The energy transfer in the process is denoted as $\hbar \omega  = E_\mathrm{in} - E_\mathrm{out}$.
The spectra were recorded from the x-ray illuminated entrance 
surface and  from the exit surface of the crystal as depicted
by the red and green lines in Fig \ref{ref:exp}).  
We repeated the experiment both off and on the 
 the Laue diffraction condition, and denote the obtained spectra 
as {\em off/on-Laue spectra} which are presented
in Fig. \ref{fig:megaplot} a)-c) and in Fig. \ref{fig:megaplot} d)-f), respectively. 
In Fig. \ref{fig:megaplot} a) we present the 
off-Laue total fluorescence yield spectrum measured from the entrance surface. 
The main edge displays a nearly indistinguishable low-energy 
shoulder at E$_{1}$=7.2405 keV, which is the quadrupole-allowed
pre-edge discussed in the introduction. 
The main edge reaches its 
maximum intensity at 7.250 keV beyond which there
is very little fine structure, and this region 
is denoted as the post edge.  
The corresponding entrance surface resonant x-ray
emission map is shown in Fig. \ref{fig:megaplot} b), and displays
distinct behaviour in the {pre-,} main- and post edge regions. 
One can clearly observe two multiplet families associated with the
2p$^5$4f$^8$5d$^0$ (pre-edge) and 2p$^5$4f$^7$5d$^1$ 
 (main edge) intermediate states, which are the most relevant for 
$L\alpha_1$ emission.
The different Coulomb interaction energies in the
final states result in an approximate 8 eV separation between these multiplets. 
The 3d$^9$4f$^8$5d$^0$ final state results in two emission peaks
separated by approximately 2 eV.  Their intensities are  maximized 
 when the incident-x-ray energy matches the pre-edge energy of 7.2405
keV. On the other hand, the   3d$^9$4f$^7$5d$^1$ peak at the high-energy
side of the emission spectrum gains its maximum when the energy of the incident 
x-rays corresponds to the main edge, i.e. 7.250 keV.
Past the main edge, the emission
spectrum is dominated by final states where the excited electron is
promoted into the continuum, resulting in a feature that disperses
linearly with the incident energy in resonant x-ray emission, leading
into off-resonant emission behaviour, i.e., x-ray fluorescence.

In Fig. \ref{fig:megaplot} c) we present the off-Laue resonant x-ray emission
spectrum recorded after tuning the 
incident x-ray energy to correspond to the pre-edge. 
The off-Laue
measurement taken from the exit surface obviously displays only background
noise, because the x-rays are strongly attenuated within the 
crystal and its back side appears as dark in the x-ray wavelengths. 
The off-Laue emission spectrum from the entrance surface, on the other hand, displays two major structures. 
A weak double peak is observed at
$\hbar\omega$=1.180--1.185 keV and asymmetric peak at 1.190 keV with
shoulders on the low and high energy sides, the former carrying more
spectral weight.
Inspection of
Fig. \ref{fig:megaplot} b) and c), and the constant final state scans
in Ref. \cite{krisch95} reveal that the intensity of the decay from the 2p$^5$5d$^1$
intermediate state exceeds the emission intensity from the 2p$^5$4f$^8$ state at all
incident x-ray energies, as expected since the latter intermediate 
state is dipole forbidden.

For the on-Laue case, the crystal is driven to the 
Laue [008] diffraction condition, and anomalous transmission is observed.
Our sample suddenly becomes semi-transparent to x rays. 
A {\em transmission} x-ray absorption spectrum can now be measured
from the diffracted beam and we show the result in Fig. \ref{fig:megaplot} d).  The quadrupole-excited 4f$^8$5d$^0$ state is
enhanced  with respect to the off-Laue case. The modification in the spectrum 
is mainly due to difference in the relative weights of E1 and E2 absorption under the 
Borrmann condition \cite{tolkiehn11}. The obtained
spectrum is overall comparable to the room temperature results of
Pettifer et al. \cite{pettifer08} However, 
the quadrupole enhancement effect on the x-ray absorption spectrum at ambient temperature is relatively modest.

Much more dramatically, the 
exit-surface on-Laue resonant x-ray emission spectra
[Fig. \ref{fig:megaplot} e)-f)] exhibits a strikingly different
behaviour from the off-Laue case.
First of all, when the anomalously
transmitted x-rays can illuminate the exit surface of the crystal, 
it becomes an strong emitter of x-rays, including L$\alpha_1$ radiation.
As an even more spectacular phenomenon, now the resonantly excited L$\alpha_1$ line shape 
has changed dramatically. The  dipole-excited 
intermediate state is diminished in spectral weight.
In contrast, the quadrupole-excited
intermediate state 4f$^8$5d$^0$ increases in relative intensity
by a factor of ten due to the Borrmann effect. 
Fig. \ref{fig:megaplot} f) shows the on-Laue resonant x-ray emission spectrum 
to be compared to the off-Laue spectra of  Fig. \ref{fig:megaplot} c).
The maximum intensity of emission from the 2p$^5$4f$^8$ intermediate
state exceeds the one originating from the 2p$^5$5d$^1$ state. This is
an important observation as the effect on x-ray absorption at room
temperature is modest in comparison with cryogenic
temperatures \cite{pettifer08,tolkiehn11}.

The suppression of the emission related 
to the dipole-allowed intermediate state from
the exit surface is not complete. We attribute the remaining
dipole related emission to lattice vibrations  
and static sample disorder (e.g., surface roughness). X-ray emission 
from the entrance surface does not contribute to the 
exit-surface signal as it would have to travel at least 15 attenuation lengths
before reaching the exit surface and the spectrometer. The on-Laue resonant X-ray emission map from
the entrance surface  was nearly identical with the one 
obtained in the off-Laue case and  is not presented. The unchanged
behaviour is to be expected on the entrance surface due to the
dominance of the dipolar absorption of the $\beta$-branch while the
$\alpha$-branch experiences reduced absorption. 

The observation of a novel type of x-ray emission excited by an anomalously 
transmitted 
x-ray beam opens up exciting possibilities to expand the capabilities 
of standing wave 
x-ray spectroscopy. Such experiments can be readily performed at synchrotron light 
sources using standard 1-eV resolution crystal spectrometers.
This will lead to a more thorough understanding of x-ray absorption spectra
and the electronic structure information derivable from the
spectrum \cite{tolkiehn11}. The observed effect could also be
utilized in precision experiments measuring the energies and 
lineshapes of weak emission features lying close in emission energy to a 
strong dipole allowed RXES channel.  
Experiments with sub-1 eV resolution on low energy
excitations, e.g. dd or crystal field excitations could exploit the
anisotropy properties discussed by Tolkiehn \textit{et al} \cite{tolkiehn11}.
For example, their experimental and
theoretical work on SrTiO$_3$ demonstrates that utilising the (110)
reflection in the perovskite structure, the quadrupolar  
transition rate to
e$_g$ states is enhanced while to the t$_{2g}$
states it is suppressed. 
Furthermore, as pointed out by Pettifer et
al. \cite{pettifer08}, the enhancement of  quadrupolar
transitions could provide a method for quantitative characterization 
of the relative weights of the multipoles 
in the x-ray absorption spectra in various systems. This would be highly useful
information in the interpretation of x-ray dichroic spectroscopies and
characterising the nature of chemical bonds in e.g. oxides and related
materials. For example, one can envision separating the effects of 
pd-hybridization and quadrupolar transitions at transition metal K edges. 
Since the Borrmann effect can be used to set up a standing wave of a desired 
periodicity and orientation, we suggest that site-selective 
spectroscopic information could be extracted via an analysis of 
standing wave measurements using appropriately chosen reflections. 

\section*{Discussion}

We have demonstrated a significant relative intensity
increase of resonant x-ray emission spectral features originating from
quadrupole transition excited intermediate states in GGG. In contrast
with the Borrmann spectroscopy introduced by Pettifer \textit{et al}.,
measuring the resonant x-ray emission spectrum under the Borrmann
condition provides a stark contrast between the 
electric dipolar and quadrupolar transitions even at room
temperature. Our results pave the way for many other interesting
studies where weak, yet-to-be-exploited features in the x-ray
absorption spectrum play a role in elucidating the electronic
structure of complex materials, e.g., when the 3d or 4f electrons
contribute to chemical bonding and magnetism. Future applications of
dynamical diffraction phenomena in spectroscopy will greatly benefit
from the improved brilliance of upcoming next generation light
sources, as improvements in beam collimation whilst maintaining 
a sufficient monochromatic photon flux on the sample will facilitate 
manipulation of
the x-ray wavefield in the sample. 
Our sample was a standard off-the-shelf commercial substrate 
and various crystals of 3d and 4f compounds of comparable quality are 
readily obtainable for further experiments. Our results
show for the first time that x-ray standing wave methods
\cite{vartanyants01} can be extended to resonant x-ray emission
spectroscopy, which is itself a widely used tool in materials,
chemistry, and physics research \cite{kotani01,glatzel05}, and thus
opens up new avenues for many research fields.

\section*{Methods}

\subsection*{Facilites}

We performed the experiment at the ID20 beamline of the European
Synchrotron Radiation Facility. We used  Si(111) crystals
to monochromatise the incident x-rays, and a crystal spectrometer 
to record the emission spectra. 
The spectrometer employed two spherically bent Si
(333) crystals in the Johann geometry 
with a bending radius of 2 m, and avalanche photodiodes as detectors.
The energy resolution was $\sim$1 eV. 

\subsection*{Experimental details}

The sample was a 150 $\mu$m thick commercial Gd$_3$Ga$_5$O$_{12}$
crystal with (001) surfaces and was kept in ambient temperature. We chose
GGG for this demonstration due to availability of high quality single crystals
and previously characterized x-ray spectra \cite{krisch95,pettifer08}. 
The sample was brought to the [008] Laue diffraction condition at the Gd L$_3$ absorption
threshold in the $\sigma$ polarisation geometry as visualized 
in Fig. \ref{ref:exp}. The incident beam  divergence was 
$<300$ $\mu$rad. The [008] reflection was chosen because it places
the nodes of the $\alpha$-branch on atomic sites occupied by the 
heavy Gd and Ga atoms \cite{pettifer08}. We verified the Borrmann effect by 
observing the transmitted and Laue-diffracted beams 
using Si $pin$ diodes behind the sample. We used similar diodes
for total fluorescence yield x-ray absorption measurements as well.
\bibliographystyle{naturemag}
\bibliography{Ruotsalainen_scirep}

\begin{thebibliography}{10}
\expandafter\ifx\csname url\endcsname\relax
  \def\url#1{\texttt{#1}}\fi
\expandafter\ifx\csname urlprefix\endcsname\relax\def\urlprefix{URL }\fi
\providecommand{\bibinfo}[2]{#2}
\providecommand{\eprint}[2][]{\url{#2}}

\bibitem{degroot01}
\bibinfo{author}{de~Groot, F.}
\newblock \bibinfo{title}{High-resolution x-ray emission and x-ray absorption
  spectroscopy}.
\newblock \emph{\bibinfo{journal}{Chemical Reviews}}
  \textbf{\bibinfo{volume}{101}}, \bibinfo{pages}{1779--1808}
  (\bibinfo{year}{2001}).
\newblock \urlprefix\url{http://dx.doi.org/10.1021/cr9900681}.

\bibitem{glatzel05}
\bibinfo{author}{Glatzel, P.} \& \bibinfo{author}{Bergmann, U.}
\newblock \bibinfo{title}{High resolution 1s core hole x-ray spectroscopy in 3d
  transition metal complexes—electronic and structural information}.
\newblock \emph{\bibinfo{journal}{Coordination Chemistry Reviews}}
  \textbf{\bibinfo{volume}{249}}, \bibinfo{pages}{65 -- 95}
  (\bibinfo{year}{2005}).
\newblock
  \urlprefix\url{http://www.sciencedirect.com/science/article/pii/S0010854504001146}.

\bibitem{carra90}
\bibinfo{author}{Carra, P.} \& \bibinfo{author}{Altarelli, M.}
\newblock \bibinfo{title}{Dichroism in the x-ray absorption spectra of
  magnetically ordered systems}.
\newblock \emph{\bibinfo{journal}{Phys. Rev. Lett.}}
  \textbf{\bibinfo{volume}{64}}, \bibinfo{pages}{1286--1288}
  (\bibinfo{year}{1990}).
\newblock \urlprefix\url{http://link.aps.org/doi/10.1103/PhysRevLett.64.1286}.

\bibitem{thole92}
\bibinfo{author}{Thole, B.~T.}, \bibinfo{author}{Carra, P.},
  \bibinfo{author}{Sette, F.} \& \bibinfo{author}{van~der Laan, G.}
\newblock \bibinfo{title}{X-ray circular dichroism as a probe of orbital
  magnetization}.
\newblock \emph{\bibinfo{journal}{Phys. Rev. Lett.}}
  \textbf{\bibinfo{volume}{68}}, \bibinfo{pages}{1943--1946}
  (\bibinfo{year}{1992}).
\newblock \urlprefix\url{http://link.aps.org/doi/10.1103/PhysRevLett.68.1943}.

\bibitem{vanderlaan99}
\bibinfo{author}{van~der Laan, G.}
\newblock \bibinfo{title}{Magnetic linear x-ray dichroism as a probe of the
  magnetocrystalline anisotropy}.
\newblock \emph{\bibinfo{journal}{Phys. Rev. Lett.}}
  \textbf{\bibinfo{volume}{82}}, \bibinfo{pages}{640--643}
  (\bibinfo{year}{1999}).
\newblock \urlprefix\url{http://link.aps.org/doi/10.1103/PhysRevLett.82.640}.

\bibitem{lucilla98}
\bibinfo{author}{Alagna, L.} \emph{et~al.}
\newblock \bibinfo{title}{X-ray natural circular dichroism}.
\newblock \emph{\bibinfo{journal}{Phys. Rev. Lett.}}
  \textbf{\bibinfo{volume}{80}}, \bibinfo{pages}{4799--4802}
  (\bibinfo{year}{1998}).
\newblock \urlprefix\url{http://link.aps.org/doi/10.1103/PhysRevLett.80.4799}.

\bibitem{kotani01}
\bibinfo{author}{Kotani, A.} \& \bibinfo{author}{Shin, S.}
\newblock \bibinfo{title}{Resonant inelastic x-ray scattering spectra for
  electrons in solids}.
\newblock \emph{\bibinfo{journal}{Rev. Mod. Phys.}}
  \textbf{\bibinfo{volume}{73}}, \bibinfo{pages}{203--246}
  (\bibinfo{year}{2001}).
\newblock \urlprefix\url{http://link.aps.org/doi/10.1103/RevModPhys.73.203}.

\bibitem{hamalainen91}
\bibinfo{author}{H\"am\"al\"ainen, K.}, \bibinfo{author}{Siddons, D.~P.},
  \bibinfo{author}{Hastings, J.~B.} \& \bibinfo{author}{Berman, L.~E.}
\newblock \bibinfo{title}{Elimination of the inner-shell lifetime broadening in
  x-ray-absorption spectroscopy}.
\newblock \emph{\bibinfo{journal}{Phys. Rev. Lett.}}
  \textbf{\bibinfo{volume}{67}}, \bibinfo{pages}{2850--2853}
  (\bibinfo{year}{1991}).
\newblock \urlprefix\url{http://link.aps.org/doi/10.1103/PhysRevLett.67.2850}.

\bibitem{vartanyants01}
\bibinfo{author}{Vartanyants, I.~A.} \& \bibinfo{author}{Kovalchuk, M.~V.}
\newblock \bibinfo{title}{Theory and applications of x-ray standing waves in
  real crystals}.
\newblock \emph{\bibinfo{journal}{Reports on Progress in Physics}}
  \textbf{\bibinfo{volume}{64}}, \bibinfo{pages}{1009} (\bibinfo{year}{2001}).
\newblock \urlprefix\url{http://stacks.iop.org/0034-4885/64/i=9/a=201}.

\bibitem{batterman64_2}
\bibinfo{author}{Batterman, B.~W.}
\newblock \bibinfo{title}{Effect of dynamical diffraction in x-ray fluorescence
  scattering}.
\newblock \emph{\bibinfo{journal}{Phys. Rev.}} \textbf{\bibinfo{volume}{133}},
  \bibinfo{pages}{A759--A764} (\bibinfo{year}{1964}).
\newblock \urlprefix\url{http://link.aps.org/doi/10.1103/PhysRev.133.A759}.

\bibitem{batterman69}
\bibinfo{author}{Batterman, B.~W.}
\newblock \bibinfo{title}{Detection of foreign atom sites by their x-ray
  fluorescence scattering}.
\newblock \emph{\bibinfo{journal}{Phys. Rev. Lett.}}
  \textbf{\bibinfo{volume}{22}}, \bibinfo{pages}{703--705}
  (\bibinfo{year}{1969}).
\newblock \urlprefix\url{http://link.aps.org/doi/10.1103/PhysRevLett.22.703}.

\bibitem{bedzyk85}
\bibinfo{author}{Bedzyk, M.~J.} \& \bibinfo{author}{Materlik, G.}
\newblock \bibinfo{title}{Two-beam dynamical diffraction solution of the phase
  problem: A determination with x-ray standing-wave fields}.
\newblock \emph{\bibinfo{journal}{Phys. Rev. B}} \textbf{\bibinfo{volume}{32}},
  \bibinfo{pages}{6456--6463} (\bibinfo{year}{1985}).
\newblock \urlprefix\url{http://link.aps.org/doi/10.1103/PhysRevB.32.6456}.

\bibitem{cowan80}
\bibinfo{author}{Cowan, P.~L.}, \bibinfo{author}{Golovchenko, J.~A.} \&
  \bibinfo{author}{Robbins, M.~F.}
\newblock \bibinfo{title}{X-ray standing waves at crystal surfaces}.
\newblock \emph{\bibinfo{journal}{Phys. Rev. Lett.}}
  \textbf{\bibinfo{volume}{44}}, \bibinfo{pages}{1680--1683}
  (\bibinfo{year}{1980}).
\newblock \urlprefix\url{http://link.aps.org/doi/10.1103/PhysRevLett.44.1680}.

\bibitem{bedzyk85_2}
\bibinfo{author}{Bedzyk, M.~J.} \& \bibinfo{author}{Materlik, G.}
\newblock \bibinfo{title}{Determination of the position and vibrational
  amplitude of an adsorbate by means of multiple-order x-ray standing-wave
  measurements}.
\newblock \emph{\bibinfo{journal}{Phys. Rev. B}} \textbf{\bibinfo{volume}{31}},
  \bibinfo{pages}{4110--4112} (\bibinfo{year}{1985}).
\newblock \urlprefix\url{http://link.aps.org/doi/10.1103/PhysRevB.31.4110}.

\bibitem{woodruff05}
\bibinfo{author}{Woodruff, D.~P.}
\newblock \bibinfo{title}{Surface structure determination using x-ray standing
  waves}.
\newblock \emph{\bibinfo{journal}{Reports on Progress in Physics}}
  \textbf{\bibinfo{volume}{68}}, \bibinfo{pages}{743} (\bibinfo{year}{2005}).
\newblock \urlprefix\url{http://stacks.iop.org/0034-4885/68/i=4/a=R01}.

\bibitem{batterman64}
\bibinfo{author}{Batterman, B.~W.} \& \bibinfo{author}{Cole, H.}
\newblock \bibinfo{title}{Dynamical diffraction of x rays by perfect crystals}.
\newblock \emph{\bibinfo{journal}{Rev. Mod. Phys.}}
  \textbf{\bibinfo{volume}{36}}, \bibinfo{pages}{681--717}
  (\bibinfo{year}{1964}).
\newblock \urlprefix\url{http://link.aps.org/doi/10.1103/RevModPhys.36.681}.

\bibitem{pettifer08}
\bibinfo{author}{Pettifer, R.~F.}, \bibinfo{author}{Collins, S.~P.} \&
  \bibinfo{author}{Laundy, D.}
\newblock \bibinfo{title}{Quadrupole transitions revealed by borrmann
  spectroscopy}.
\newblock \emph{\bibinfo{journal}{Nature}} \textbf{\bibinfo{volume}{454}},
  \bibinfo{pages}{196--199} (\bibinfo{year}{2008}).
\newblock \urlprefix\url{http://dx.doi.org/10.1038/nature07099}.

\bibitem{carra91}
\bibinfo{author}{Carra, P.}, \bibinfo{author}{Harmon, B.~N.},
  \bibinfo{author}{Thole, B.~T.}, \bibinfo{author}{Altarelli, M.} \&
  \bibinfo{author}{Sawatzky, G.~A.}
\newblock \bibinfo{title}{Magnetic x-ray dichroism in gadolinium metal}.
\newblock \emph{\bibinfo{journal}{Phys. Rev. Lett.}}
  \textbf{\bibinfo{volume}{66}}, \bibinfo{pages}{2495--2498}
  (\bibinfo{year}{1991}).
\newblock \urlprefix\url{http://link.aps.org/doi/10.1103/PhysRevLett.66.2495}.

\bibitem{lang92}
\bibinfo{author}{Lang, J.~C.} \emph{et~al.}
\newblock \bibinfo{title}{Circular magnetic x-ray dichroism at the erbium
  ${\mathit{l}}_{3}$ edge}.
\newblock \emph{\bibinfo{journal}{Phys. Rev. B}} \textbf{\bibinfo{volume}{46}},
  \bibinfo{pages}{5298--5302} (\bibinfo{year}{1992}).
\newblock \urlprefix\url{http://link.aps.org/doi/10.1103/PhysRevB.46.5298}.

\bibitem{wang93}
\bibinfo{author}{Wang, X.}, \bibinfo{author}{Leung, T.~C.},
  \bibinfo{author}{Harmon, B.~N.} \& \bibinfo{author}{Carra, P.}
\newblock \bibinfo{title}{Circular magnetic x-ray dichroism in the heavy
  rare-earth metals}.
\newblock \emph{\bibinfo{journal}{Phys. Rev. B}} \textbf{\bibinfo{volume}{47}},
  \bibinfo{pages}{9087--9090} (\bibinfo{year}{1993}).
\newblock \urlprefix\url{http://link.aps.org/doi/10.1103/PhysRevB.47.9087}.

\bibitem{lang95}
\bibinfo{author}{Lang, J.~C.} \emph{et~al.}
\newblock \bibinfo{title}{Confirmation of quadrupolar transitions in circular
  magnetic x-ray dichroism at the dysprosium ${\mathit{l}}_{\mathrm{iii}}$
  edge}.
\newblock \emph{\bibinfo{journal}{Phys. Rev. Lett.}}
  \textbf{\bibinfo{volume}{74}}, \bibinfo{pages}{4935--4938}
  (\bibinfo{year}{1995}).
\newblock \urlprefix\url{http://link.aps.org/doi/10.1103/PhysRevLett.74.4935}.

\bibitem{giorgetti95}
\bibinfo{author}{Giorgetti, C.} \emph{et~al.}
\newblock \bibinfo{title}{Quadrupolar effect in x-ray magnetic circular
  dichroism}.
\newblock \emph{\bibinfo{journal}{Phys. Rev. Lett.}}
  \textbf{\bibinfo{volume}{75}}, \bibinfo{pages}{3186--3189}
  (\bibinfo{year}{1995}).
\newblock \urlprefix\url{http://link.aps.org/doi/10.1103/PhysRevLett.75.3186}.

\bibitem{krisch95}
\bibinfo{author}{Krisch, M.~H.} \emph{et~al.}
\newblock \bibinfo{title}{Evidence for a quadrupolar excitation channel at the
  ${\mathit{l}}_{\mathrm{iii}}$ edge of gadolinium by resonant inelastic x-ray
  scattering}.
\newblock \emph{\bibinfo{journal}{Phys. Rev. Lett.}}
  \textbf{\bibinfo{volume}{74}}, \bibinfo{pages}{4931--4934}
  (\bibinfo{year}{1995}).
\newblock \urlprefix\url{http://link.aps.org/doi/10.1103/PhysRevLett.74.4931}.

\bibitem{tolkiehn11}
\bibinfo{author}{Tolkiehn, M.}, \bibinfo{author}{Laurus, T.} \&
  \bibinfo{author}{Collins, S.~P.}
\newblock \bibinfo{title}{Effects of temperature and anisotropy on quadrupole
  absorption in borrmann spectroscopy}.
\newblock \emph{\bibinfo{journal}{Phys. Rev. B}} \textbf{\bibinfo{volume}{84}},
  \bibinfo{pages}{241101} (\bibinfo{year}{2011}).
\newblock \urlprefix\url{http://link.aps.org/doi/10.1103/PhysRevB.84.241101}.

\end{thebibliography}

\section*{Acknowledgements}

We acknowledge the European Synchrotron Radiation Facility for beamtime.  
This work was supported by Academy of Finland projects 1254065, 1283136, 1259526, 1260204 and 1259599.

\section*{Author contributions statement}

K.R. concieved the experiment, K.R., A-P.H., S.C., G.M, M.M., M.K, and S.H. planned and 
performed the experiment, K.R. and S.H. analysed the experimental data.  
All authors contributed to and reviewed the manuscript. 

\section*{Competing financial interests}
The authors declare no competing financial interests.
\end{document}